\documentclass{article}
\usepackage{multirow}
\usepackage{adjustbox}
\usepackage{spconf,amsmath,graphicx,color, amssymb}
\usepackage{booktabs}
\usepackage{makecell}
\usepackage{array}
\usepackage{caption}
\usepackage{subcaption}  % 使用subcaption宏包进行子图排列
\usepackage{tabularx}

\usepackage{url}
\title{SELM: Speech Enhancement Using Discrete Tokens and Language Models}
\name{Ziqian Wang$^1$, Xinfa Zhu$^1$, Zihan Zhang$^1$, YuanJun Lv$^1$, Ning Jiang$^2$, Guoqing Zhao$^2$, Lei Xie$^1$$^{*}$\thanks{* Corresponding author.}}
\address{
$^1$Audio, Speech and Language Processing Group (ASLP@NPU), School of Computer Science, \\Northwestern Polytechnical University, Xian, China\\
$^2$Mashang Consumer Finance Co., Ltd.}
\begin{document}
\ninept
\maketitle

\begin{abstract}

% 语言模型在其他领域中表现优异，展示出了强大的语义理解能力。对于SE，理解语义会更有利于这个任务
%Deep learning-based speech enhancement has made significant progress, while the use of Large Language Models (LMs) in this domain remains unexplored. Conventional algorithms rely on continuous speech representations, hindering the effective utilization of discrete token-based LMs. To address this issue,

Language models (LMs) have recently shown superior performances in various speech generation tasks, demonstrating their powerful ability for semantic context modeling. Given the intrinsic similarity between speech generation and speech enhancement, harnessing semantic information is advantageous for speech enhancement tasks. In light of this, we propose SELM, a novel speech enhancement paradigm that integrates discrete tokens and leverages language models. SELM comprises three stages: encoding, modeling, and decoding. We transform continuous waveform signals into discrete tokens using pre-trained self-supervised learning (SSL) models and a k-means tokenizer. Language models then capture comprehensive contextual information within these tokens. Finally, a de-tokenizer and HiFi-GAN restore them into enhanced speech. Experimental results demonstrate that SELM achieves comparable performance in objective metrics and superior subjective perception results. %Moreover, SELM exhibits the added benefit of enhancing speech recognition accuracy.
Our demos are available \footnote{\url{https://honee-w.github.io/SELM/}}.

%Moreover, SELM exhibits promising generalization ability, effectively %handling noise types unseen during training.
%\vspace{-3pt}

\end{abstract}
%\renewcommand{\thefootnote}{\fnsymbol{footnote}}
%\footnotetext{* Corresponding author.}
%
\vspace{-10pt}
\section{Introduction}
\vspace{-5pt}
\label{sec:intro}
 % - 现有的方式及优缺点
Speech enhancement (SE) is a fundamental task in speech processing, aiming to improve the quality and intelligibility of speech signals corrupted by various noise sources. Over the last few years, deep learning techniques have demonstrated remarkable advancements in SE, enabling more effective noise reduction. %These techniques can be broadly categorized into two groups: discriminative and generative approaches.%
    
%Deep learning-based SE can be roughly divided into two categories based on the criteria used to estimate the transformation from noisy speech to clean speech. The first category trains \emph{discriminative models} to minimize the distance between noisy and clean speech. However, as supervised methods are inevitably trained on a finite set of training data with limited model capacity for practical reasons, they may not generalize well to unseen situations, e.g., different noise types, different signal-to-noise ratios (SNR), and reverberations. Moreover, some of these methods have been observed to introduce undesirable speech distortions~\cite{wang2019bridging}. The second category trains \emph{generative models} to learn the distribution of clean speech as a prior for speech enhancement instead of learning a direct noisy-to-clean mapping. Several approaches have employed deep generative models for speech enhancement, including generative adversarial networks (GANs)~\cite{pascual2017segan}, variational autoencoders (VAEs)~\cite{fang2021variational, bie2022unsupervised}, flow-based models~\cite{nugraha2020flow}, and diffusion probabilistic models~\cite{lu2021study,lu2022conditional}. Existing studies~\cite{fang2021variational, bie2022unsupervised, lu2022conditional} have shown better performance of generative SE on unseen testing noises than discriminative counterparts, making it a promising direction for improving generalization capability.

%第一句感觉很不通顺
Deep learning-based speech enhancement can be roughly divided into two categories, based on the criteria used to estimate the transformation from noisy speech to clean speech.
The first category uses \emph{discriminative models} to minimize the distance between noisy and clean speech. However, as supervised methods are inevitably trained on a finite set of data, they may not generalize well to unseen situations, such as different noise types and reverberations. Some of these methods can even introduce unwanted speech distortions~\cite{wang2019bridging}. The second category employs \emph{generative models}, which learn the distribution of clean speech as a prior for enhancement rather than directly mapping from noisy speech to clean speech. Within this category, various deep generative models, including generative adversarial networks (GANs)~\cite{pascual2017segan}, variational autoencoders (VAEs)~\cite{fang2021variational, bie2022unsupervised}, flow-based models~\cite{nugraha2020flow}, and diffusion probabilistic models~\cite{lu2022conditional, lemercier2023storm, sgmse}, have been used for SE. Studies~\cite{fang2021variational, bie2022unsupervised, lu2022conditional} have demonstrated that generative methods perform better on unseen noises than discriminative methods, making it a promising direction for improving generalization capability.

%In this context, we investigate the application of Language Large Models (LMs) to speech enhancement tasks, leveraging their robust generative capabilities. Recently, LMs have emerged as powerful tools in natural language processing, exhibiting remarkable performance across various tasks~\cite{zhao2023survey}. These models are typically based on transformer architectures~\cite{vaswani2017attention}, pre-trained on massive language datasets, and have the ability to capture intricate linguistic patterns and context. As a result of their success in natural language processing, there has been growing interest in exploring the applications of LMs in other domains, particularly in speech-related tasks such as automatic speech recognition~\cite{radford2023robust} and speech synthesis~\cite{wang2023neural} due to the fact that speech shares many similarities with natural language. The key advantage of using LMs in these tasks lies in their ability to model complex contextual relationships within speech data effectively.  However, the potential of LMs in speech enhancement remains largely unexplored due to the challenge of integrating continuous speech representations with discrete token-based LMs. Furthermore, existing speech enhancement methods using continuous representations may introduce unnecessary computations and hinder the utilization of large datasets, affecting the model's capabilities and generalization ability.

Self-supervised learning (SSL) has significantly impacted the field of speech processing by enabling unsupervised learning from large amounts of unlabeled speech data~\cite{chen2022wavlm}. These SSL features have proven highly effective in various downstream tasks~\cite{yang21c_interspeech, song2023exploring}. Notably, the discretization of these representations allows for generating high-quality speech tokens from continuous speech, creating new opportunities in spoken language modeling~\cite{lakhotia2021generative}.

Recent advancements in leveraging LMs with discrete tokens for various speech generation tasks~\cite{wang2023neural, wu2023speechgen, wang2023speechx}, such as speech synthesis, continuation, and impainting, have gained significant attention. Correspondingly, the inherent similarities between speech enhancement and speech generation have led to efforts to apply LM to speech front-end processing~\cite{erdogan2023tokensplit, le2023voicebox}. These systems often rely on transcribed text to facilitate speech enhancement or separation and underperform without linguistic aids from the transcripts. However, a significant challenge arises in real-world scenarios where transcriptions are either limited or entirely absent. This calls for a comprehensive exploration of how to integrate LMs into speech enhancement effectively. Furthermore, given that LMs normally rely on discrete tokens, how to address the issue of preserving intricate details like speech attributes, speaker characteristics, and background sounds during the transformation from continuous representations to discrete tokens also demands further investigation.

To address these limitations, we propose a novel framework named SELM, inspired by Vec-Tok Speech~\cite{vectokspeech}, 
which effectively integrates LMs by introducing discrete tokens that encapsulate both acoustic and semantic information. This bridging of continuous and discrete representations allows us to harness the full potential of LMs for contextual modeling in speech enhancement tasks. We adapt the mask prediction paradigm of LMs to the SE task, treating it as a specialized form of mask prediction where the model learns to transform noisy speech into clean speech by minimizing the difference between predicted and target tokens. Our framework consists of three key stages: encoding, modeling, and decoding. Specifically, in the \emph{encoding} stage, we transform raw audio waveforms into discrete tokens using pre-trained WavLM~\cite{chen2022wavlm} and k-means clustering~\cite{macqueen1967some}, competently capturing and preserving the intricate and diverse information in the continuous speech signals, facilitating efficient signal reconstruction. Subsequently, in the \emph{modeling} stage, we leverage language models to capture comprehensive contextual information within these discrete tokens, allowing for effective modeling of complex relationships and dependencies. Finally, in the \emph{decoding} stage, we integrate a de-tokenizer and HiFi-GAN~\cite{Kong2020HiFiGANGA} to restore these tokens into continuous speech, ensuring the enhanced speech retains natural and coherent characteristics. In summary, the main contributions of this work lie in the exploration of LMs for speech enhancement and the development of SELM, a framework that leverages language models to enhance speech signals effectively. Through extensive experiments, we demonstrate the advantages of SELM over existing approaches. SELM achieves comparable performance in objective metrics while outperforming in subjective perception. %Additionally, SELM demonstrates improvements in speech recognition accuracy. 
%and exhibits promising generalization ability to unseen noise types during training.
    
%By introducing LMs to the domain of speech enhancement, we aim to demonstrate their potential and encourage further progress in speech-related applications. The knowledge obtained from this study suggests possible benefits for practical situations, especially in improving speech communication in noisy settings and contributing to the enhancement of speech-processing systems.%

%\vspace{-10pt}
\section{PROPOSED APPROACH}
%\vspace{-8pt}

\iffalse
\subsection{Preliminary}
In real environments, received noisy signal $x(t)$ in the time domain can be modeled as:
\begin{equation}
    x(t) = s(t) * h(t) + n(t) ,
\end{equation}
where $*$ denotes the convolution operator, $s(t)$, $h(t)$, and $n(t)$ denote clean signal, room impulse response (RIR), and background noise, respectively. For legibility, we will omit the subscript if no conflict arises.

Speech enhancement is a task to restore high-quality clean speech $\hat{s}$ from $x$:
\begin{equation}
      \hat{s} = e(x),    
\end{equation}
where $e$ is the enhancement function, whose target is to estimate $s$ by restoring $\hat{s}$ from the received signal $x$.
In this study, we set $e$ as the proposed SELM,  which aims to simultaneously achieve denoising, dereverberation, and restoration. A diagram of the framework can be seen in Figure \ref{fig:selm}.
\fi 

\subsection{Framework Overview}
\begin{figure*}[]
% \vspace{-8pt}
  \centering
  \includegraphics[width=0.9\linewidth]{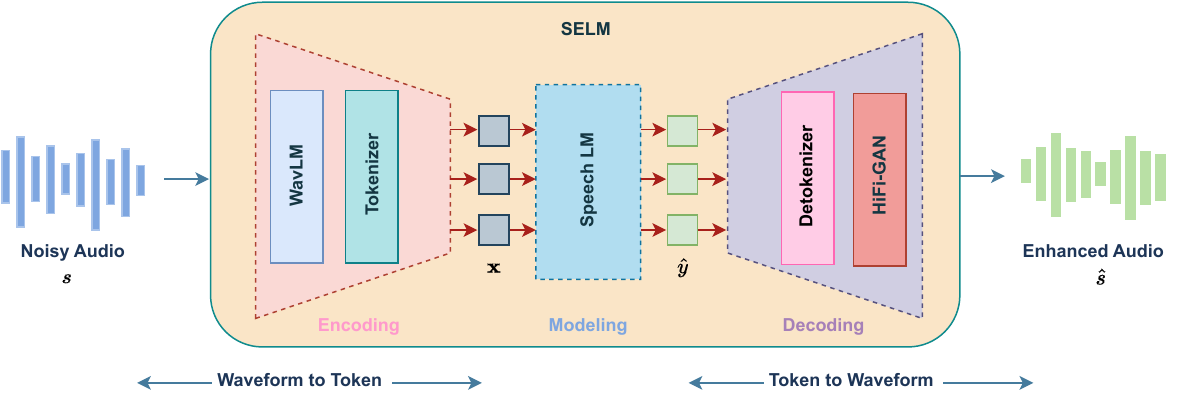}
\caption{Overview of the SELM framework. SELM handles SE tasks by using transformer-based LMs conditioned on discrete tokens.}
\label{fig:selm}
\vspace{-4pt}
\end{figure*}

\vspace{-2pt}
We present SELM, a novel framework that leverages LMs using discrete tokens for speech enhancement, as shown in Figure \ref{fig:selm}. SELM comprises three key components: a speech encoder, a speech language model (LM), and a speech decoder. %Initially, the speech encoder takes the input corrupted waveform $s \in  \mathbb{R}^{T}$ and transforms it into a token sequence $\mathbf{x} = \left( x_1, \dots, x_{T\prime }\right) $. Next, the speech LM operates as a language model for token sequences, optimizing the likelihood by predicting the upcoming token $\hat{y}_t$ based on the prior tokens $\hat{y}_{<t}$ and the token sequence $\mathbf{x}$. Finally, the speech decoder processes the tokens generated by the speech LM to convert them back into the enhanced waveform $\hat{s}$. The pipeline can be formulated as follows:
The speech encoder, which includes WavLM and k-means tokenizer, initially transforms the input corrupted waveform $s \in \mathbb{R}^{T}$ into a token sequence $\mathbf{x} = \left( x_1, \dots, x_{T\prime }\right) $. Subsequently, the speech LM operates as a language model for token sequences, optimizing the likelihood by predicting the upcoming token $\hat{y}_t$ based on the prior tokens $\hat{y}_{<t}$ and the token sequence $\mathbf{x}$. The speech decoder, comprising a de-tokenizer and HiFi-GAN, finally processes the tokens generated by the speech LM to recover them back into the enhanced waveform $\hat{s}$.  The pipeline can be formulated as follows:
\begin{align}
% \footnotesize
\mathbf{x} & = \text{Tokenizer}\left(\text{WavLM}(s)\right), \label{eq:line1} \\
\hat{\mathbf{y}} & = \text{SpeechLM}\left(\textbf{x}\right), \label{eq:line2} \\
\hat{s} & = \text{HiFi-GAN}\left(\text{Detokenizer}\left(\hat{\mathbf{y}}\right)\right). \label{eq:line3}
\vspace{-6pt}
\end{align}
%where Tokenizer maps each embedding to the index of the closest k-means codebook entry~\cite{borsos2023audiolm}, detokenizer transforms predicted tokens into enhanced audio waveforms.

\vspace{-4pt}
\subsection{Speech Encoder}

In our SELM framework, the speech encoder plays a pivotal role in both feature extraction and tokenization. Leveraging a pre-trained SSL model, specifically WavLM Large~\cite{chen2022wavlm}, we harness its unique ability to capture intricate acoustic and semantic information. This fusion of a CNN encoder and a Transformer empowers the WavLM model with a comprehensive grasp of contextual cues, facilitated by masking during training. Given a corrupted signal $s$, we employ WavLM to extract continuous representations $r$. Notably, we select representations from %the 6th layer of%
WavLM as input features for incorporating both acoustic and semantic information, making it more suitable for waveform reconstruction~\cite{song2023exploring}.  %based on our observation. 
Subsequently, the k-means tokenizer maps continuous speech features $r$ to discrete tokens $\textbf{x}$. By calculating and updating centroids, each frame's representation $r_{t} \in \mathbb{R}^{F}$ are transformed into a scalar value $\in \left( 0, K-1\right)$, representing $K$ distinct clusters~\cite{borsos2023audiolm}, where $F$ denotes the feature dimension. 
\iffalse
The encoding stage can be outlined as Eq.(\ref{eq:encoder}), where $T$, $F$, and $T^{\prime}$ denote sampling points, feature dimensions, and time frames, respectively:

\begin{equation}
\label{eq:encoder}
 s \in \mathbb{R}^T  \longrightarrow  r \in \mathbb{R}^{F \times T'}  \longrightarrow d \in \mathbb{R}^{T'}
\end{equation}
\fi 
%\[
%\begin{array}{ccl}
%    x \in \mathbb{R}^T & \longrightarrow &  f \in \mathbb{R}^{F \times T'} \\
%    f \in \mathbb{R}^{F \times T'} & \longrightarrow & d \in \mathbb{R}^{T'}
%\end{array}
%\]

\vspace{-4pt}
\subsection{Speech LM}
\vspace{-2pt}
LMs have demonstrated remarkable performance across diverse downstream tasks without needing task-specific training from scratch. The efficacy of LMs is largely attributed to the pivotal technique of mask prediction~\cite{radford2018improving}, which plays a crucial role in their pre-training phase. By exposing the model to extensive data, the model is trained to predict masked tokens within a sequence. This process equips the model with a deep comprehension of linguistic structures and semantics, enhancing its ability to grasp and apply language patterns.

In our approach, we draw inspiration from this mask prediction paradigm and apply it to the SE task. We view the SE task as a specialized form of mask prediction based on our observation that noisy speech tokens and their corresponding clean speech tokens share substantial similarities. This resemblance allows us to perceive the distinct portions as indicative of distortions, effectively translating into masked areas. Consequently, the speech LM is fed with noisy speech tokens while utilizing tokens from corresponding clean speech as learning targets. By minimizing the discrepancy between the predicted tokens and the target tokens, the model is forced to learn the transformation from noisy to clean speech and comprehend contextual information.

We follow the decoder-only fashion as GPT-series models~\cite{radford2018improving, radford2019language, brown2020language}, utilizing the Transformer architecture~\cite{vaswani2017attention} as the backbone of the speech LM for its excellent parallelizability and capacity. Speech LM in SELM predicts the discrete tokens required to synthesize enhanced speech. The attention mechanism is applied to the input tokens, considering the entire set of tokens to calculate attention scores and enabling bidirectional modeling to capture comprehensive contextual information. This allows the model to learn acoustic and linguistic details, including speaker characteristics and understanding the context within the tokens.
 In language modeling, the speech LM aims to predict the target probability distribution based on the input probability distribution.

\begin{figure}[]
% \vspace{-8pt}
  \centering
  \includegraphics[height=0.75\linewidth]{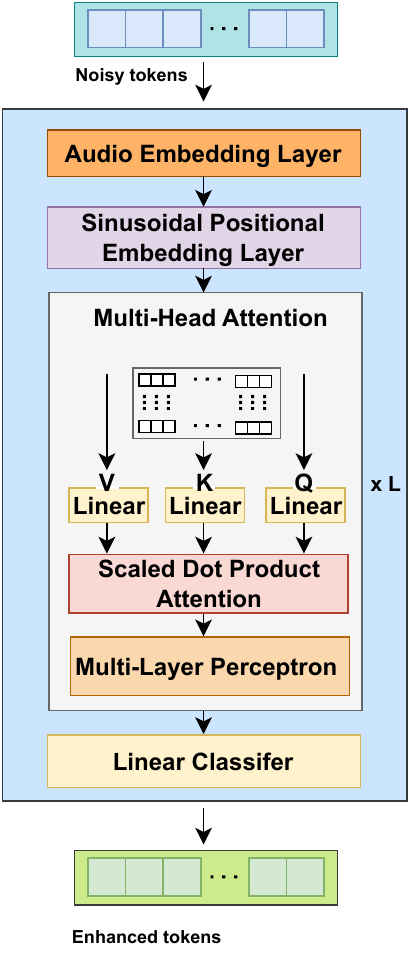}
\caption{The details of the modeling stage.}
\label{fig:LM}
\vspace{-8pt}
\end{figure}

To achieve this, corrupted token sequences $\textbf{x}$ are first passed through an audio embedding layer, producing continuous output representations. To encode positional information, sinusoidal positional embeddings are incorporated into the token embeddings. The resulting sequences are fed into the transformer decoder, consisting of stacked multi-head self-attention layers and MLP blocks. At each layer, layer normalization is applied before and after, and a residual connection is used. The MLP consists of two linear fully-connected layers with a GELU non-linear activation function. Once we obtain the learned representations, we pass them to a Linear Classifier to predict the enhanced token sequences $\hat{\textbf{y}}$. A detailed pipeline of the modeling stage is shown in Figure \ref{fig:LM}.

In this modeling approach, our primary learning objective is to train the speech LM to predict the target probability distribution based on the input probability distribution. Denote $\textbf{p}(\textbf{y})$ as the probability distribution over target tokens, and $\textbf{q}(\hat{\textbf{y}})$ as the probability distribution over predicted tokens. The learning objective can be formulated as shown in Eq (\ref{eq:loss_for_lm}):

\begin{equation}
% \footnotesize
\label{eq:loss_for_lm}
\mathcal{L}_{\text{speechLM}} = -\sum^{T^\prime}_{i=1}p\left(\textbf{y}_{i} \right) \dot{} \text{log} \left( \frac{q(\hat{\textbf{y}}_i)}{p(\textbf{y}_{i})} \right).  
\vspace{-4pt}
\end{equation} 

\vspace{-4pt}
\subsection{Speech Decoder}
\vspace{-2pt}

Within the Speech Decoder phase, a de-tokenizer and HiFi-GAN~\cite{Kong2020HiFiGANGA} are facilitated to reconstruct refined audio signals $\hat{s}$ from enhanced tokens $y$. The former, comprising a token embedding layer and a conformer decoder~\cite{gulati2020conformer}, effectively translates discrete tokens back into fine-grained continuous representations $\hat{r}$. Specifically, a lookup table is maintained as the token embedding layer, associating discrete tokens' values $\in (0, K-1)$ with the k-means tokenizer's clustering center vectors.  The conformer decoder captures contextual nuances, strengthened by a mask prediction strategy. The learning targets of the de-tokenizer  are the continuous representations of the corresponding target speech obtained by WavLM Large, as shown in the following formulations:
\begin{align}
r_{target} &= \text{WavLM}(s_{target}), \\
\mathcal{L_{\text{detokenizer}}} &=   \frac{\sum^{T^\prime}_{i}(r_{target_i}-\hat{r}_i)^2}{T^\prime}. 
\end{align}
%\begin{equation}\
%\label{eq:loss_for_detokenizer}
%        \mathcal{L}_{\text{detokenizer}} = \text{MSELoss}(r, \text{(WavLM}(s))
%\end{equation}
%MSE loss is used to optimize the prediction of SSL model representations $\hat{f}$.%
The transformed continuous representations $\hat{r}$ then pass through HiFi-GAN. In our approach, HiFi-GAN processes the predicted WavLM features instead of the conventional Mel spectrum conditioning, transiting $\hat{r}$ to enhanced audio signals $\hat{s}$.

\iffalse
The decoding stage can be formulated as:
\begin{equation}
    \hat{d} \in \mathbb{R}^{T^{\prime}} \longrightarrow  \hat{r} \in \mathbb{R}^{F \times T^{\prime}} \longrightarrow \hat{s} \in \mathbb{R}^{T}
\end{equation}
\fi
\vspace{-8pt}
\section{EXPERIMENTS}
\label{sec:typestyle}

\vspace{-4pt}
\subsection{Datasets and evaluation metrics}
\vspace{-2pt}
\iffalse
\textbf{Training sets} The training speech datasets we use include a subset of WeNetSpeech%~\cite{Yao2021WeNetPO}
, GigaSpeech%~\cite{chen2021gigaspeech}
, LibriMix%~\cite{cosentino2020librimix}
, VoiceBank, and the DNS challenge - Interspeech 2021 datasets~\cite{reddy21_interspeech}. The noise datasets are a subset of WHAM!~\cite{wichern2019wham} and DEMAND~\cite{thiemann2013diverse}. To simulate the reverberations, room impulse responses (RIRs), which are from
the openSLR26 and openSLR28~\cite{ko2017study}, are randomly chosen. All training data are generated on the fly with an 80\% probability of adding noise recordings with a signal-to-noise ratio (SNR) drawn from a Gaussian distribution with $N$(−5, 20) dB and a 50\% probability to be convolved with a piece of room impulse response.
\fi 

\textbf{Training sets} The speech encoder and decoder are trained on a subset of WeNetSpeech~\cite{Yao2021WeNetPO} and GigaSpeech~\cite{chen2021gigaspeech}, 2000 hours of Mandarin and English speech. As for the speech LM, 
The training speech datasets we use include a subset of LibriMix~\cite{cosentino2020librimix}
, VoiceBank (VCTK)~\cite{voicebank}, and the DNS challenge - Interspeech 2021 datasets~\cite{reddy21_interspeech}. The noise datasets are a subset of WHAM!~\cite{wichern2019wham} 
and DEMAND~\cite{thiemann2013diverse}. 
Room impulse responses (RIRs), which are from
the openSLR26 and openSLR28~\cite{ko2017study}, are randomly chosen to simulate the reverberations. All training data are generated on the fly with an 80\% probability of adding noise recordings with a signal-to-noise ratio (SNR) from $\mathcal{N} $(\text{-}5, 20) dB and a 50\% probability to be convolved with room impulse responses.

\textbf{Test sets} We utilize the publicly available DNS challenge - Interspeech 2021 test set to compare our framework with existing state-of-the-art models. 
 %The efficacy of the various models is assessed using this test set.  
 In addition,  we simulate noisy mixtures with speech samples randomly picked from VCTK and noise and RIRs unseen during training to measure SELM's improvement in automatic speech recognition (ASR) accuracy.

\textbf{Evaluation metrics} Since the output of the vocoder is not strictly aligned on the sample level with the target, as is often the case in generation models~\cite{kumar2020nugan}, this effect will degrade the metrics, especially for those calculated on time samples such as Si-SNR. Therefore, the objective metrics we adopt include DNSMOS and Log Likelihood Ratio (LLR). DNSMOS~\cite{reddy2021dnsmos} is a neural network-based mean opinion score (MOS) estimator that was shown to correlate well with the quality ratings by humans. LLR is an LPC-based objective measure with a high correlation with overall quality~\cite{semetrics}. We use the pre-trained ECAPA-TDNN~\cite{Desplanques2020ECAPATDNNEC} to extract the x-vector and calculate the cosine similarity to verify the speaker similarity.
Inspired by ITU-T P.835, we further conduct mean opinion score (MOS) experiments to evaluate speech quality (QMOS) and speaker similarity (SMOS). Specifically, we randomly select 30 testing samples from each test set for subjective evaluations involving a group of 15 listeners. Moreover, word error rate (WER) is adopted to measure the ASR accuracy and intelligibility improvements.%\footnote{\url{https://github.com/wenet-e2e/wenet/tree/main/examples/librispeech/s0}}. 
 %We use mean opinion scores (MOS) to evaluate different systems subjectively. 

\begin{table*}[htbp]
    \caption{The performance of different systems on the DNS Challenge testset. ``G'' and ``D'' denote generative and discriminative categories.}
    \centering
    \resizebox{\textwidth}{!}{
    \begin{tabular}{@{}ccccccccccccccc@{}}
    \toprule
    \multirow{3}{*}{System} &
      \multirow{3}{*}{Category} &
      \multicolumn{5}{c}{With Reverb} &
      \multicolumn{5}{c}{Without Reverb} &
      \multicolumn{3}{c}{Real Recordings} \\
    \cmidrule(lr){3-7} \cmidrule(lr){8-12} \cmidrule(lr){13-15}
     & & \multicolumn{3}{c}{DNSMOS $\uparrow$} & \multirow{2}{*}{LLR $\downarrow$} & \multirow{2}{*}{Cosine Sim. $\uparrow$} & \multicolumn{3}{c}{DNSMOS $\uparrow$} & \multirow{2}{*}{LLR $\downarrow$} & \multirow{2}{*}{Cosine Sim. $\uparrow$} & \multicolumn{3}{c}{DNSMOS $\uparrow$} \\
    \cmidrule(lr){3-5} \cmidrule(lr){8-10} \cmidrule(lr){13-15}
     & & SIG & BAK & OVL & & & SIG & BAK & OVL & & & SIG & BAK & OVL \\
    \midrule
    \addlinespace[1ex]
    Noisy & - & 1.760 & 1.497 & 1.392 & - & - & 3.392 & 2.618 & 2.483 & - & - & 3.053 & 2.510 & 2.255 \\
    \midrule
    \addlinespace[1ex]
    Conv-TasNet & D & 2.415 & 2.710 & 2.010 & 0.283 & 0.939 & 3.092 & 3.341 & 3.001 & 0.241 & 0.945 & 3.102 & 2.975 & 2.410 \\
    Demucs & D & 2.510 & 2.641 & 2.215 & \textbf{0.201} & \textbf{0.941} & 3.124 & 3.257 & 3.010 & 0.235 & 0.952 & 2.974 & 2.870 & 2.291 \\
    Inter-SubNet & D & 2.651 & 2.581 & 2.362 & 0.248 & 0.933 & 3.458 & 3.821 & 3.099 & \textbf{0.207} & \textbf{0.967} & 3.258 & 3.574 & 2.806 \\
    \midrule
    \addlinespace[1ex]
    CDiffuSE & G & 2.541 & 2.300 & 2.190 & 0.320 & 0.907 & 3.294 & 3.641 & 3.047 & 0.317 & 0.914 & 3.201 & 3.104 & 2.781 \\
    SGMSE & G & 2.730 & 2.741 & 2.430 & 0.314 & 0.899 & 3.501 & 3.710 & 3.137 & 0.310 & 0.934 & 3.297 & 2.894 & 2.793 \\
    StoRM & G & 2.947 & 3.141 & 2.516 & 0.303 & 0.934 & \textbf{3.514} & 3.941 & 3.205 & 0.299 & 0.943 & 3.410 & 3.379 & 2.940 \\
    \midrule
    \addlinespace[1ex]
    \hspace{1em} SELM($K$=300) & G & \textbf{3.160} & \textbf{3.577} & \textbf{2.695} & 0.334 & 0.901 & 3.508 & \textbf{4.096} & \textbf{3.258} & 0.320 & 0.917 & \textbf{3.591} & \textbf{3.435} & \textbf{3.124} \\
    \hspace{1em} SELM($K$=1024) & G & 3.001 & 3.471 & 2.574 & 0.317 & 0.923 & 3.417 & 3.871 & 3.126 & 0.303 & 0.941 & 3.430 & 3.246 & 3.005 \\
    \hspace{1em} SELM($K$=4096) & G & 2.991 & 3.214 & 2.490 & 0.310 & 0.940 & 3.497 & 3.901 & 3.214 & 0.293 & 0.950 & 3.464 & 3.137 & 3.037 \\
    \bottomrule
    \end{tabular}
    }
    \label{tab:results}
\end{table*}

\vspace{-6pt}
\subsection{Training setup and baselines}
\vspace{-2pt}
We utilize pre\text{-}trained WavLM Large and select the embeddings of the 6th layer for k\text{-}means clustering. The number of clusters $K$ is set to 300, 1024, and 4096, respectively. The speech LM consists of an audio embedding layer, a sinusoidal positional embedding, and 12 transformer decoder blocks with a hidden dimension of 1024 and 16 attention heads. The dimension of the de-tokenizer is determined by the number of clusters.  We follow the official settings of HiFi\text{-}GAN and substitute spectrograms with predicted WavLM features as input. Speech LM and the detokenizer are trained using 8 NVIDIA 3090 GPUs with a batch size of 24 per GPU for 80 epochs. We use the AdamW optimizer with a learning rate of $5\times 10^{-4}$ for speech LM and $1 \times 10^{-4}$ for the de-tokenizer. Exponential decay updates the learning rate after each epoch, using a decay ratio $1 \times 10^{-5}$.

 We compare the performance of our proposed method with 3 generative and 3 discriminative baselines, which we describe in more detail below. All models are re-trained by us following the official implementations with the same training data.
\begin{itemize}
\vspace{-2pt}
\item Conv-TasNet~\cite{luo2019conv}: An end-to-end neural network that estimates a filtering mask.

\item Demucs~\cite{defossez2020real}: A U-Net-based discriminative model that operates in the waveform domain.

\item Inter-Subnet~\cite{chen2023inter}: A subband-based discriminative model with subband interaction.

\item SGMSE~\cite{sgmse}: A score-based generative model for speech enhancement using a deep complex U-Net.

\item CDiffuSE~\cite{lu2022conditional}: A generative speech enhancement method based on a conditional diffusion process in the time domain.

\item StoRM~\cite{lemercier2023storm}: A score-based generative model for speech enhancement with the stochastic regeneration scheme.
\end{itemize}\

\vspace{-8pt}
\subsection{Results and discussions}
\vspace{-2pt}
%\subsubsection{Comparison with competitive baselines}
Table \ref{tab:results} shows the performance of different systems on the DNS Challenge test dataset. Over the table, ``With Reverb", ``Without Reverb", and ``Real Recordings" stands for the paired synthetic test sets with and without reverberation, as well as the unlabeled real recordings test set, respectively. ``$K$" represents the clustering number in the k-means tokenizer of SELM. DNSMOS is calculated for all test sets as it is a reference-free metric; LLR and cosine similarity are calculated for the synthetic test sets with clean speech labels.

%The performance of different SELM variants is compared using the last three rows of Table \ref{tab:results}. Increasing cluster numbers improves DNSMOS metrics but slightly decreases LLR and cosine similarity measures. In other words, fewer clusters yield better objective metrics, while more clusters enhance subjective perception. A possible reason is that fewer clusters result in lower learning difficulty and better convergence, leading to improved objective outcomes. Conversely, more clusters heighten learning complexity, increasing inference errors. However, greater clustering allows for more fine-grained modeling, enabling better capture of the information related to timbre, emotions, pitch, and other aspects of original noisy mixture, thereby enhancing subjective perception.

In Table \ref{tab:results}, we assess different SELM variants by examining the last three rows. 
%We find a trade-off between cluster numbers and performance metrics. When the number of clusters increases, DNSMOS metrics improve, but LLR and cosine similarity measures slightly decrease. 
We find a trade-off between cluster numbers and performance metrics. Decreasing $K$ improves DNSMOS metrics while marginally reducing LLR and cosine similarity. In essence, fewer clusters yield superior objective metrics due to easier learning and better convergence, improving objective metrics. Conversely, more clusters increase learning complexity but enable finer modeling of timbre, emotions, pitch, and other aspects in the noisy mixture, thus enhancing subjective perception. To optimize the clustering process, we conduct additional experiments to reduce the number of clusters. We observe that setting $K$ to less than 300 leads to poor intelligibility and subjective perception. Consequently, these results are not included in the Table \ref{tab:results}.

Moreover, we can draw a comparison between discriminative and generative systems. Notably, the generative models generally perform better in perception-related metrics such as DNSMOS due to the ability to alleviate the problems of noise leakage and excessive suppression to a certain extent. However, due to the output of the generative model not necessarily being strictly consistent with the target at the sample level,
%because the output of the generated model may not necessarily correspond strictly to the target at the sample level, 
they may underperform in the LLR and cosine similarity metrics.

We extend the comparison of our proposed SELM with state-of-the-art discriminative and generative methods using the DNS Challenge dataset in Table \ref{tab:results}. The conclusion can be drawn that the SELM shows superb performance in the denoising task. This indicates that discrete representation holds promise for addressing speech enhancement challenges. Additionally, the language modeling capability of LM enhances the effectiveness of generative models, and the impressive performance of objective metrics validates the efficacy of our approach.

%\vspace{-10pt}
%\subsubsection{Improvments for human and machine perception}

%The results in Table \ref{wer} highlight the ability of our proposed SELM model, which achieves the best performance on the simulated testset. Notably, SELM's robust generalization ability to unseen noise during training, along with its effectiveness in enhancing perceptual quality for both humans and machines, underscores its potential for real-world applications.

%Additionally, We visually analyze the unprocessed and processed signals in the simulated testset as shown in Fig \ref{fig:sample_spec}. It can be observed that there is still obvious noise leakage in the noisy mixture after processingrou through Inter-SubNet. Conversely, signals enhanced by SELM exhibit high similarity to the target clean speech, effectively reducing noise without introducing additional artifacts.

Table \ref{tab:perception} presents the performances of different systems on the simulated test set. WER is measured for the source noisy audio and the processed audio by an ASR model\footnote{\url{https://github.com/wenet-e2e/wenet/tree/main/examples/librispeech/s0}}, indicating speech intelligibility. Subjective listening tests are conducted to obtain QMOS and SMOS, respectively, reflecting speech quality and speaker similarity. Comparing SELM to the comparative baseline systems, several observations can be made. In terms of WER, SELM outperforms all benchmark systems. This demonstrates that SELM significantly enhances speech intelligibility, effectively preserving the linguistic information of the source audio. While SELM performs slightly lower than the top-performing system regarding speaker similarity (SMOS), it's essential to note that SELM maintains competitive performance in generative models. Furthermore, SELM excels in terms of speech quality. This showcases SELM's capability to generate high-quality audio signals, making speech more natural and pleasant to the listener. Our demos are available at https://honee-w.github.io/SELM/.

\begin{table}[]
\caption{WER and MOS scores with 95\% confidence intervals on the simulated testset}
\centering
\resizebox{0.95\linewidth}{!}{
\begin{tabular}{ccccc}
\hline
System       & Category & WER $\downarrow$           & QMOS $\uparrow$         & SMOS $\uparrow$         \\ \hline
Noisy        & $-$        & 10.87         & 2.70 $\pm$ 0.08         & $-$             \\ \hline
Conv-TasNet  & D        & 7.63          & 3.00 $\pm$ 0.07       & 3.93 $\pm$ 0.09         \\
Demucs       & D        & 7.01          & 3.07 $\pm$0.10        & 3.98 $\pm$ 0.11         \\
Inter-SubNet & D        & 5.03          & 3.24 $\pm$ 0.09        & \textbf{4.01 $\pm$ 0.07} \\ \hline
CDiffuSE     & G        & 8.14          & 3.13 $\pm$ 0.08       & 3.88 $\pm$ 0.10         \\
SGMSE        & G        & 7.45          & 3.27 $\pm$ 0.07        & 3.90 $\pm$ 0.10        \\
StoRM        & G        & 6.14          & 3.37 $\pm$ 0.10       & 3.97 $\pm$ 0.08        \\ \hline
SELM         & G        & \textbf{4.32} & \textbf{3.43 $\pm$ 0.08} & 3.95 $\pm$ 0.07         \\ \hline
\end{tabular}
}
    \label{tab:perception}
\end{table}

\iffalse
\begin{figure}[htbp]
    \centering
    \begin{minipage}{0.22\linewidth}
        \centering
        \includegraphics[width=\linewidth]{clean.pdf}
    \end{minipage}\hfill
    \begin{minipage}{0.22\linewidth}
        \centering
        \includegraphics[width=\linewidth]{mix.pdf}
    \end{minipage}\hfill
    \begin{minipage}{0.22\linewidth}
        \centering
        \includegraphics[width=\linewidth]{intersubnet.pdf}
    \end{minipage}\hfill
    \begin{minipage}{0.22\linewidth}
        \centering
        \includegraphics[width=\linewidth]{gen_2.pdf}
    \end{minipage}
    \caption{Mel spectrograms of clean, noisy, Inter-SubNet enhanced and SELM enhanced signals, respectively.}
    \label{fig:sample_spec}
\end{figure}
\fi

\vspace{-6pt}
\section{Conclusions}
\vspace{-2pt}
In this paper, we introduced SELM, a novel paradigm for speech enhancement that combines discrete tokens and Language Models (LMs). Pretrained SSL models and a k-means tokenizer enable LMs to model contextual information effectively. The de-tokenizer and HiFi-GAN recover the enhanced speech signals further. Experimental results reveal SELM's comparable performance in objective metrics and superior performance in subjective perception. We hope our work inspires future research on using LMs for speech enhancement and related signal improvement tasks.

\vfill\pagebreak
\clearpage
% \section{REFERENCES}
% \label{sec:refs}

% List and number all bibliographical references at the end of the
% paper. The references can be numbered in alphabetic order or in
% order of appearance in the document. When referring to them in
% the text, type the corresponding reference number in square
% brackets as shown at the end of this sentence ~\cite{C2}. An
% additional final page (the fifth page, in most cases) is
% allowed, but must contain only references to the prior
% literature.

% \bibliographystyle{IEEEtran}
\footnotesize
\bibliographystyle{IEEE}
\bibliography{strings,refs}

\end{document}